\documentclass[preprint, authoryear, 5p, number]{elsarticle}
\usepackage{hyperref}
\usepackage{units}
\usepackage{amssymb}
\usepackage[english]{babel}
\usepackage{color}
\usepackage{array}
\usepackage{booktabs}
\usepackage{graphicx}
\usepackage{array}
\usepackage{subfig}
\usepackage{fnpos}

\usepackage[T1]{fontenc}
\hypersetup{
    colorlinks=true,
    linkcolor=blue,
    filecolor=magenta,      
    urlcolor=blue,
}

\newcolumntype{M}[1]{>{\centering\arraybackslash}m{#1}}
\newcolumntype{N}{@{}m{0pt}@{}}

\journal{Astronomy and Computing}

\begin{document}

\begin{frontmatter}

\title{FRELLED : A Realtime Volumetric Data Viewer For Astronomers}

\author[a]{R. Taylor\corref{cor1}}
\ead{rhysyt@gmail.com}
\cortext[cor1]{Corresponding author}
\address[a]{Astronomical Institute of the Czech Academy of Sciences, Prague}

\begin{abstract}
I present a new FITS viewer designed to explore 3D spectral line data (in particular H\,\textsc{i}) and assist with visual source extraction and analysis. Using the artistic software Blender, \textsc{frelled} can visualise even large ($\sim$600$^{3}$ voxels) data sets at high frame rates ($\gtrsim$10 f.p.s.) in 3D. Blender's interface enables easy navigation within the 3D environment, and the \textsc{frelled} scripts support world coordinate systems. A variety of tools are included to aid source extraction and analysis, including interactively masking data (using 3D polyhedra of arbitrary complexity), querying NED, calculating the flux in specified volumes, generating contour plots and overlaying optical data. It includes tools to overlay n-body particle data, and multi-volume rendering is supported. The interface is designed to make cataloguing sources as easy as possible and I show that this can be as much as a factor of 50 times faster than using other viewers.
\end{abstract}


\end{frontmatter}

\section{Introduction}
Searching 3D data cubes is not a straightforward task. It is non-trivial to design a source-finding algorithm which is flux-limited, and complex, spurious structures can be difficult to automatically distinguish from astronomical sources. While visual inspection of the data can, through the extremely powerful capabilities of human pattern recognition, often overcome the problem of distinguishing real from spurious sources, it suffers from being subjective. It is also, traditionally, an extremely slow and laborious process, to the point where the scientific returns are diminished : weeks (or even months) spent searching a data cube could be more usefully employed in analysing trends in the data.

I here focus on neutral hydrogen (H\,\textsc{i}) data cubes. Developments in multibeam single-dish surveys (e.g. HIPASS, \citealt{barnes}; ALFALFA, \citealt{aa}) have made very large 3D data sets a reality. Smaller, more sensitive surveys such as AGES (the Arecibo Galaxy Environment Survey - \citealt{auld}) possess similar problems of source extraction. Although the total number of sources is lower, the source density is higher, as shown in table \ref{tab:surveys}. This is necessarily a crude comparison (since not every survey provides access to the raw data, accurately estimating what fraction of the data the sources occupy is impossible), but suffices to demonstrate the very substantial increase in the discovery potential of H\,\textsc{i} surveys over the last decade or so.

\begin{table*}
\begin{center}
\caption[HI data]{Notable HI surveys and their source densities. Columns : (1) Survey name; (2) Total area in square degrees; (3) Velocity range in km\,s$^{-1}$; (4) Total number of sources detected; (5) Number of sources detected per unit volume relative to HIPASS. HIPASS data is from \cite{meyer}, HIJASS from \cite{lang}, HIDEEP from \cite{hideep}, ALFALFA from \cite{a40}, and AGES from AGES papers I-VIII (all published data at the time of writing).}
\label{tab:surveys}
\begin{tabular}{c c c c c}
\hline
  \multicolumn{1}{c}{(1)Survey} &
  \multicolumn{1}{c}{(2)Area} &
  \multicolumn{1}{c}{(3)Velocity range / km s$^{-1}$} &
  \multicolumn{1}{c}{(4)Sources} &
  \multicolumn{1}{c}{(5)Sources/volume}\\
\hline
  HIPASS & 34,432 & -1,280 $<$ v $<$ 12,700 & 4,315 & 1.0\\
  HIJASS & 1,115 & -3,500 $<$ v $<$ 10,000 & 222 & 1.6\\
  HIDEEP & 60 & -1,280 $<$ v $<$ 12,700 & 173 & 23.0\\
  ALFALFA & 2,800 & -2,000 $<$ v $<$ 20,000 & 15,855 & 28.7\\
  AGES & 65 & -2,000 $<$ v $<$ 20,000 & 886 & 69.1\\
\hline
\end{tabular}
\end{center}
\end{table*}

Problems for visual source extraction in recent H\,\textsc{i} surveys are primarily due to two reasons : (1) The sheer number of sources. When viewing a data cube, not even an experienced observer can possibly remember the locations of hundreds or thousands of detections; (2) Source density. The higher the fraction of data which contains emission (be it from astrophysical sources or artificial Radio Frequency Interference), the more difficult it is for an observer to remember which source they have catalogued - even if the total number of sources is low.

These difficulties will only increase as new telescopes come online and existing facilities are upgraded. \cite{P15}, hereafter P15, have summarised the capabilities of some of the anticipated projects at the ASKAP and MeerKAT telescopes, as well as the greatly enhanced capabilities of the Westerbork interferometer offered by the installation of the APERTIF instrument\footnote{I only add here that Arecibo is also expected to receive a similar upgrade, the AO40 phased array instrument, which would increase survey speeds by approximately a factor of four (\citealt{g08}).}. Importantly, while the existing surveys take years and produce catalogues of a few thousand sources, the next generation surveys are expected to detect thousands of sources \textit{per week}.

From the perspective of source extraction, the challenges of these extremely large data sets are formidable. One solution would be increased reliance on catalogues generated by purely automatic algorithms. In some ways, this is the ideal solution : automatic source extractors are objective and give repeatable results. In practise, there are many reasons that it is best to avoid relying on such algorithms exclusively given the present state of the art. Not least of these is that even for detecting point sources, their reliability is very poor at low S/N levels (reliability here meaning what fraction were found to be real astrophysical sources with follow-up observations).

\cite{me13} investigated the reliability of several source-finding programs at the 4$\sigma$ level using an AGES data cube. We found a reliability level of 2\% for \textsc{polyfind} (\citealt{d01}; \textsc{multifind} has a similar reliability level - \citealt{meyer}), 7\% for \textsc{duchamp} (\citealt{duch}, designed to work with any 3D FITS data) and 16\% for \textsc{glados} (\citealt{me12}, specifically written to deal with AGES data). Reliability of both  \textsc{glados} and \textsc{duchamp} can be increased by a factor of two but at the expense of completeness (see \citealt{mythesis}). However, note that the volumes searched by the automatic routines had to be specially chosen to exclude regions affected by radio frequency interference (RFI) - without this restriction the reliability levels drop by at least an order of magnitude. In contrast, the reliability of a visual extraction catalogue was estimated at 51\% and included regions strongly affected by RFI.

Even if automatic source-finding routines can be improved, there are other reasons why visual extraction is desirable. Unforeseen problems with the data (or, equally, real sources with unexpected characteristics) may confuse automatic programs  which a human could rapidly identify and understand. An intuitive feeling for how a source relates to others in the data set can be invaluable for understanding unusual characteristics (e.g. intergalactic bridges and tails). Additionally, large surveys are only part of the story - a detailed understanding of the data usually relies on extracting smaller catalogues focused on particular objects, and knowledge of the individual sources becomes essential.

In this paper I present \textsc{frelled}, the FITS Realtime Explorer of Low Latency in Every Dimension. \textsc{Frelled} is designed to address one of the main disadvantages to visual source extraction : its low speed. Human pattern recognition is able to understand massively more complex images than those of H\,\textsc{i} data cubes on very short ($<<$ 1 second) time scales - it is not a problem of finding the sources, but \textit{recording} them. \textsc{Frelled} provides tools that overcome this problem, allowing the user to record hundreds of sources per day. While many FITS viewers allow 3D views of the data as a secondary feature, with \textsc{frelled} this is an intrinsic part of the process. It thus provides a new way to explore and understand 3D data sets.

The rest of this paper is organised as follows. Section \ref{sec:code} discusses the code and its performance in comparison to other viewers (focusing on the display capabilities). Section \ref{sec:astrofrelled} gives an overview of the features useful for astronomical data analysis in \textsc{frelled}, comparing them with other tools. In section \ref{sec:future} I discuss some of the shortcomings of \textsc{frelled} and possible future improvements, and I summarise the discussion in section \ref{sec:summary}.

\textsc{Frelled} relies on the artistic software `Blender'. Throughout this article I will refer to Blender when dealing with a feature explicitly provided by the software with no Python code required, and \textsc{frelled} when referring to features which are enabled - or made simpler - by my scripts. A complete archive of all \textsc{frelled} releases, including source code, can be found \href{http://www.rhysy.net/frelled-archive.html}{online}\footnote{\href{http://www.rhysy.net/frelled-archive.html}{http://www.rhysy.net/frelled-archive.html}}. A detailed user guide is also available on the \textsc{frelled} \href{http://frelled.wikia.com/wiki/FRELLED\_Wikia}{wiki page}
\footnote{\href{http://frelled.wikia.com/wiki/FRELLED\_Wikia}{http://frelled.wikia.com/wiki/FRELLED\_Wikia}}.

\section{Overview of \textsc{frelled}'s display capabilities}
\label{sec:code}
\subsection{Blender's display interface}
There were two primary motivations for developing \textsc{frelled} : (1) a desire to view 3D data sets in 3D, thus enabling a rapid, intuitive understanding of the data; (2) the need to be able to easily, rapidly, and interactively mask regions of the data (for instance to hide RFI and record galaxies).

To accomplish this task, I use the open-source software Blender. While Blender itself is not designed as a data viewer (it is primarily aimed at artists), it has many characteristics that make it eminently suitable as such. The most important of these for astronomy is the presence of a Python interpreter. This allows Blender to interface with any of the numerous astronomical routines available in Python. Secondly, it has a powerful interactive 3D display - both artists and astronomers have a mutual interest in viewing large amounts of data. The advantage of using Blender is that to leapfrog the need to develop a display routine and concentrate on the features required for astronomy (the disadvantage is being at the mercy of the Blender developers regarding display capabilities). For a thorough review of the capabilities of Blender in astronomy, see \cite{bkent}.

In brief, Blender is designed for creating 3D content (still images, videos, and interactive features). The primary way a user does this is by generating an object which consists of a mesh of vertices that define virtual faces. A vertex is simply a set of x,y,z coordinates. A face is a plane defined by three or four vertices. The mesh is the set of vertices and their associated faces. Meshes can be assigned different materials to give colour and transparency, which may include the use of textures (procedurally-generated functions or pre-computed images). Each mesh is part of an object, which can have position, scale, and rotation along the X/Y/Z axes. Each object contains a single mesh, but different objects may contain the same mesh. Most parameters can be animated (e.g. position, material colour, rotation, etc.) - that is, made to vary in a controlled way over some specified number of frames.

\subsection{Displaying 3D FITS data into Blender via \textsc{frelled}}
\label{sec:displaycode}
\textsc{Frelled} is a set of Python scripts which can be used to visualise 3D astronomical FITS files inside Blender. On a typical system (at least 2 GB RAM, a 1 GB GPU and a 1.5 GHz CPU), Blender is capable of handling a few ($<$ 10) million vertices at $\sim$ 25 f.p.s., but only a few thousand objects (which contain much more data than vertices, and so require much more memory). In order to display the full FITS data for a 500 voxel$^{3}$ cube using its internal mesh format, Blender would require at least one vertex and one material per pixel (i.e. 125 million). This is far in excess of what Blender is designed to handle, and consequently it is not a practical solution - frame rates would drop well below 1 f.p.s. Thus unfortunately we cannot, for example, simply write a Python script to load in every voxel as a cube and give it some transparency, which might be the most obvious solution \footnote{For an experiment in using this method, including the Python code, see \href{http://www.rhysy.net/fits-method-2-1.html}{this link : http://www.rhysy.net/fits-method-2-1.html}.}. This method would require much of the faint emission to be clipped, thus precluding it as a suitable approach for source finding (which must preserve as much of the original data as possible).

This limitation can be circumvented using the technique known as \textit{billboarding}. First, the cube is ``sliced'' into a series of PNG images, each of a constant velocity, declination, or right ascension. The images are generated using a combination of the standard \textit{pyfits} and \textit{matplotlib} Python modules.

\begin{figure}[t]
\begin{center}
\includegraphics[width=84mm]{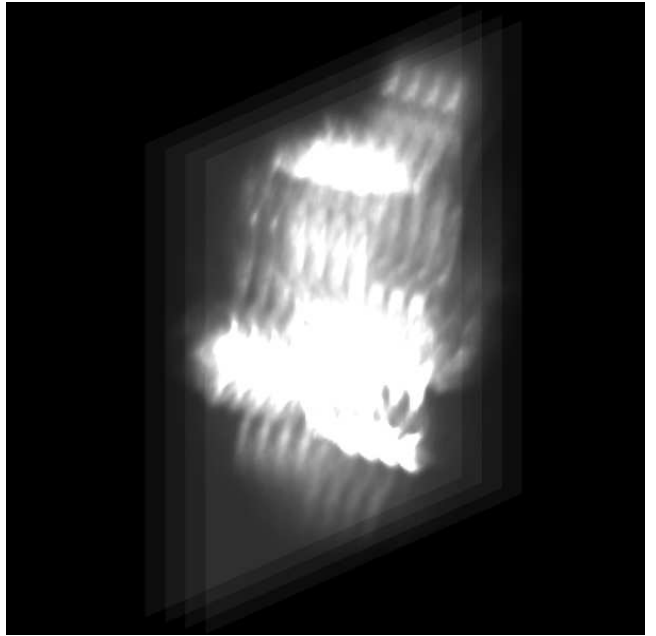}
\caption[Bill]{Simple example of the billboarding technique. Here four image planes are used, each with the same image texture. To illustrate the technique more clearly, a minimum level of transparency has been set so that the planes themselves are visible, and only a single image texture is used. For displaying volumetric data, typically tens or hundreds of image planes are used, each with an image texture from a different slice of the data. With low data levels allowed to be completely transparent, this can be an effective way to display volumetric data.}
\label{fig:bill}
\end{center}
\end{figure}

These images are mapped onto a sequence of Blender's internal mesh objects (planes), which are separated by an amount equivalent to the size of one pixel in the PNG images. A simple example of this is shown in figure \ref{fig:bill}. Blender is quite capable of handling hundreds of objects with image textures at $>$ 25 f.p.s., so we are able to view the full information of the FITS file at a high frame rate. To view the data in 3D, the materials are given some transparency which is influenced by the image texture - i.e. any region above some threshold will be completely opaque, and any below another threshold will be completely transparent. This lower limit means that some of the data is clipped out, but typically far less than if each voxel were a separate cube object. If the data has had good baseline subtraction (i.e. noise distribution symmetrical about a value of 0.0), clipping all values below 0.0 is normally sufficient.

While the level of clipping required to display data in 3D in \textsc{frelled} is minimal, it is not zero. The realtime display engine of Blender calculates the transparency as the sum of the values along the line of sight. This means that in 3D mode some faint sources can be missed. Therefore \textsc{frelled} also incorporates a conventional 2D mode, which I will briefly discuss in section \ref{sec:2D}. I will mainly concentrate on the more novel 3D aspect of the program in the rest of the paper.

Note that \textsc{frelled} uses the PNG images only for displaying the data inside Blender. All analysis tasks which require access to the original data (which I discuss in section \ref{sec:astrofrelled}) use the original FITS file.

As evident in figure \ref{fig:bill}, displaying the data on a series of planes has the obvious disadvantage that those planes cannot be seen if the view is directly edge-on. This can be avoided by slicing the data along the three major axes (or `projections', as in the \textit{kvis} parlance; \citealt{karma}). Following the \textit{kvis} convention we term these the XY (normally sky coordinates), XZ (RA-velocity) and ZY (Dec-velocity) projections. Displaying each projection requires a separate collection of image planes. The user can control which projections they want to load into \textsc{frelled}, reducing load time and memory use.

To save further memory and increase the realtime display frame rate, only one projection is displayed at a time (each projection is stored on a separate Blender ``layer'' or workspace). A Python script continuously monitors the angle of the viewpoint and switches to the appropriate projection in realtime. If the user only loaded in one or two projections, the display will only show the view appropriate for those projections. Thus the display is optimised to save time, memory, and preserve the best possible display of the data. I discuss this further in section \ref{sec:opt}.

It is possible to use different colour \textit{transfer functions} to control the transparency and colour of the data. We caution that this is a completely different concept to the \textit{radiative} transfer function, which may be more familar to users with a background in astrophysical theory. The radiative transfer function determines how light propogates through a medium and the wavelength-dependent changes dues to scattering, absorption and emission. This is useful in visualisations, for example, in converting the results of numerical simulations into maps of emission in different wavebands (e.g \citealt{lagos}). However this is far from a trivial procedure and not implemented in \textsc{frelled} (or indeed most FITS viewers).

\begin{figure*}[t]
\begin{center}
\includegraphics[width=180mm]{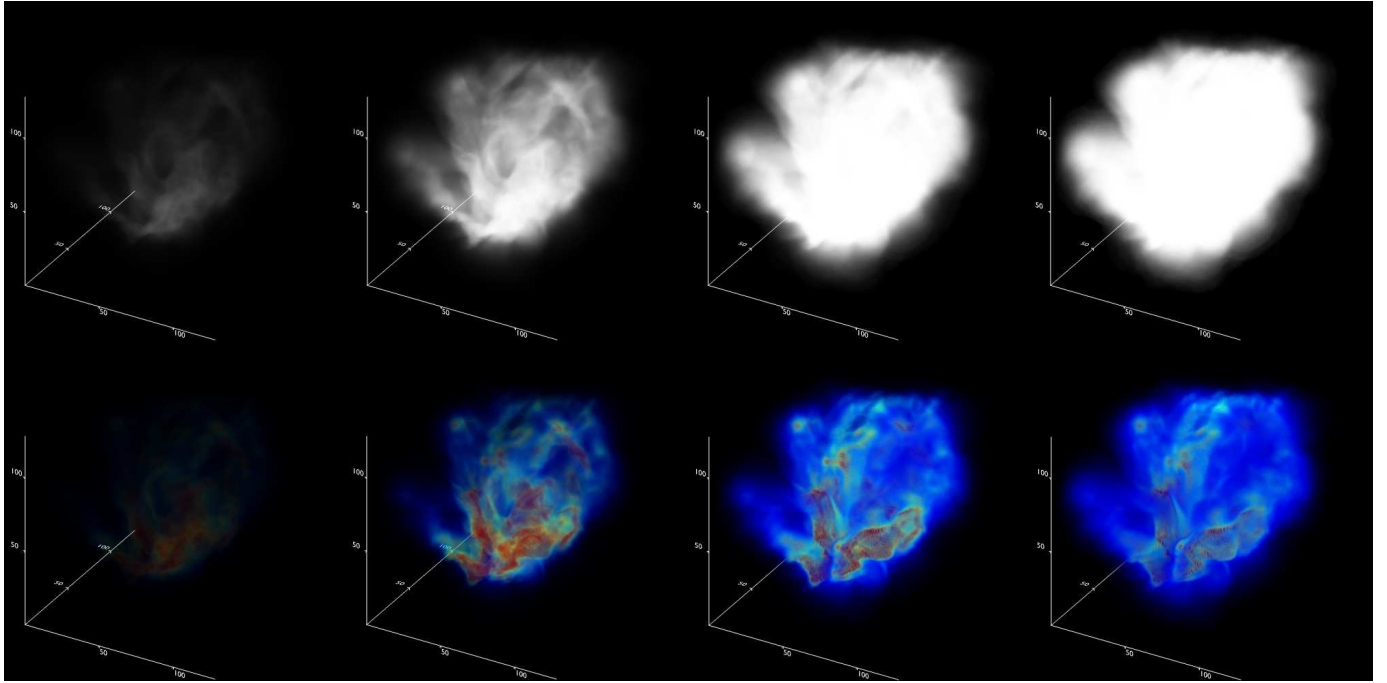}
\caption[display]{Examples of using different display settings in \textsc{frelled} to reveal different features in a (numerical simulation) data cube. The top row shows the cube displayed using a single greyscale colour map to affect only the opacity. The second row shows the use of a combination of a greyscale opacity map and a rainbow colour map. From left to right, the maximum opacity (1.0 = opaque) varies from 0.01, 0.1, 0.5, to 1.0.}
\label{fig:display}
\end{center}
\end{figure*}

In contrast the term ``transfer function'' in computer graphics normally refers to the \textit{colour} transfer function, which is a much simpler concept (this is what we will refer to throughout the remainder of the paper). The colour transfer function determines which RGB values (and in the case of \textsc{frelled} also the transparency, alpha) to assign a pixel based on the input data values. Mathematically this can be trivial. For example, a data value of 0.0 might be assigned RGBA values of 0.0, displaying as transparent black, while data values of 1.0 might be given RGBA values of 1.0, shown as opaque white, with a linear interpolation of values in between (the standard greyscale mapping). Far more complex mappings of the data to RGBA values are also possible : the RGB components do not have to have the same gradient or vary monotonically (a simple example is shown in figure \ref{fig:display}).

\textsc{Frelled} includes standard linear and logarithmic transfer scaling functions found in many other astronomical FITS viewers (see section \ref{sec:others}). This is implemented via the \textit{matplotlib} Python module, along with a variety of different colour schemes provided as standard in \textit{matplotlib}\footnote{See the following url :\\ \href{http://matplotlib.org/examples/color/colormaps_reference.html}{http://matplotlib.org/examples/color/colormaps\_reference.html}}. This module also allows users to create their own transfer functions which may be more suitable to their particular data set (we discuss some common problems specific to astronomy in more detail in section \ref{sec:astrodata}). To do this the user has only to specify the RGBA values they wish at specified data values, with interpolation between the missing values being linear (see, for example, \href{http://matplotlib.org/examples/pylab\_examples/custom\_cmap.html}{this url)}\footnote{\href{http://matplotlib.org/examples/pylab\_examples/custom\_cmap.html}{http://matplotlib.org/examples/pylab\_examples/custom\_cmap.html}}.

For transparency, Blender does an on-the-fly conversion of RGB to alpha values (by averaging). Therefore it is much easier to understand a greyscale map for controlling the transparency and to use another function to determine the colour. \textsc{Frelled} allows the user to use two different maps, with one or both affecting the colour and transparency.

The ability to use different transfer functions enables an interesting possibility : one can use one FITS file to set the transparency and other data set to set the colour. This may be particularly useful for simulations, where one has access not only to density information at every point but also (for example) temperature and velocity. It also has potential applications for multi-wavelength observational data (see section \ref{sec:multi}). 

\subsection{Astronomical data}
\label{sec:astrodata}
Astronomical data presents a specific set of challenges. To some extent, the data sets are so diverse and specialised that processing the data to display it in the ``best'' way will have to be the responsibility of the user. For example, due to the hexagonal pattern of the multibeam receiver, AGES data cubes have higher noise levels at the spatial edges. Without correcting for this the higher flux levels at the edge cause the edges of the data to appear opaque in \textsc{frelled}, blocking the view of most of the data. In \cite{me14} (hereafter T14) we described how recasting the cube in signal to noise (dividing each pixel value by the \textit{rms} along that spectrum) can remove this effect without losing any visual information (see T14 appendix A). I cannot possibly hope to incorporate all such incredibly specialised solutions within \textsc{frelled} (however the code for this particular operation is available \href{http://www.rhysy.net/Resources/Programs/FitsBaseline_SN.py}{online}\footnote{\href{http://www.rhysy.net/Resources/Programs/FitsBaseline_SN.py}{http://www.rhysy.net/Resources/Programs/FitsBaseline\_SN.py}}) - the user will always have to do \textit{some} work to exploit their data to its full potential.

\begin{figure*}[t]
\begin{center}
\includegraphics[width=180mm]{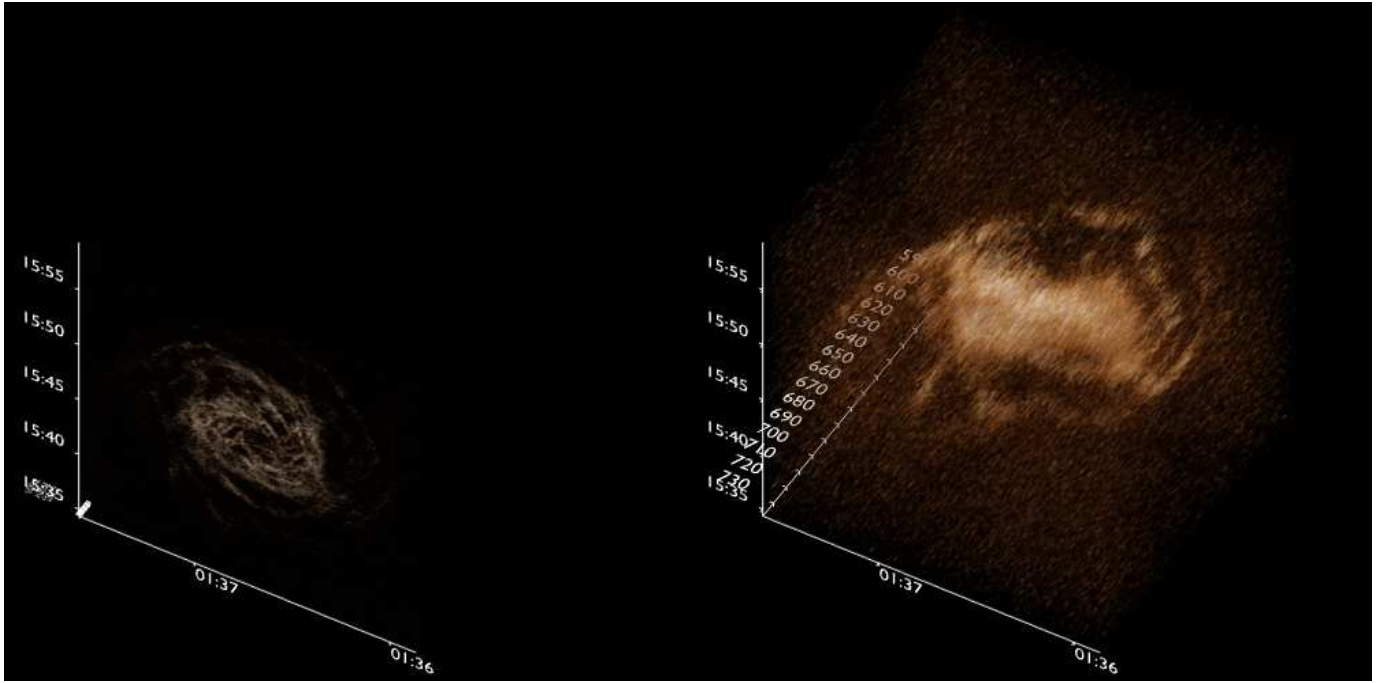}
\caption[stretch]{NGC 628 H\,\textsc{i} data cube (from The H\,\textsc{i} Nearby Galaxy Survey, \citealt{things}) as displayed in \textsc{frelled}. The left panel shows the data with every voxel at the same size - since the number of channels (58) is low compared with the number of spatial pixels (1024), kinematic information is essentially lost in the display. The right panel shows the data with the data stretched along the velocity axis so as to be displayed as a cube - kinematic information is readily apparent. The same transfer function is used in both cases.}
\label{fig:stretch}
\end{center}
\end{figure*}

Nonetheless, there are some common aspects to most astronomical data that for which some solutions can be anticipated. The third axis is usually velocity (redshift), not distance, and the number of channels depends on velocity resolution and bandwidth (i.e. the receiver). This means that the size of one channel has no correlation whatsoever with the size of one spatial pixel - indeed they contain qualitatively different information. When the number of channels is low compared to the number of spatial pixels, displaying them with equal size can make velocity information much harder to discern (and yet strangely other astronomical 3D viewers, such as \textit{d9} and \textit{xray}, do not account for this). For this reason, \textsc{frelled} includes an option to stretch the velocity range of the displayed data to be equal to the spatial size (x-axis, usually right ascension) of the data. For an example, see figure \ref{fig:stretch}.

Data often has a high dynamic range, and a logarithmic intensity scale can be as useful in revealing faint features in 3D as it can in 2D. However unlike 2D data, in 3D data cubes the dynamic range can vary significantly from channel to channel. A transfer function which displays the full range of data in one channel might show nothing at all in others. This problem is particularly acute for Galactic H\,\textsc{i} data. Simply using a linear or logarithmic transfer function over some data range results in either displaying the high-flux channels clearly, but rendering the low-flux channels invisible, or showing the low-flux channels clearly but rendering the high-flux channels opaque.

\begin{figure*}[t]
\begin{center}
\includegraphics[width=180mm]{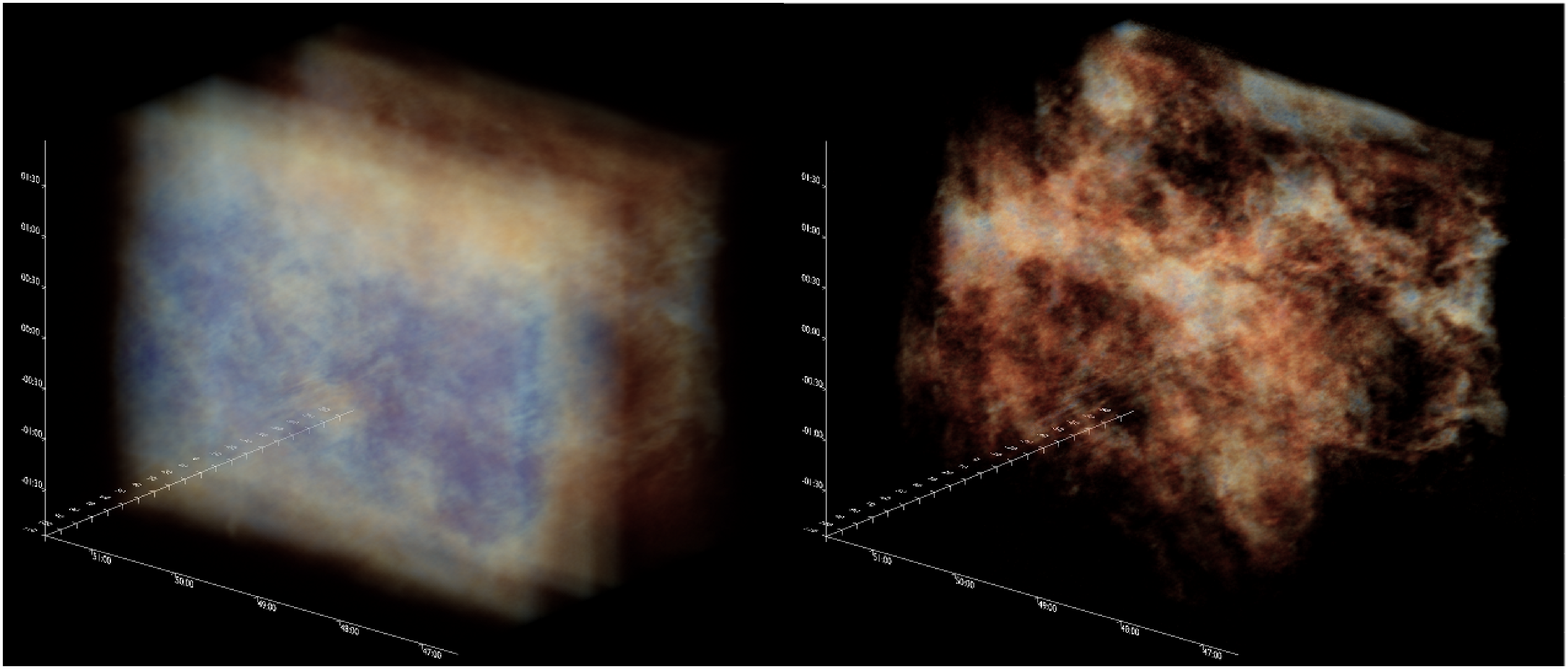}
\caption[scale]{VLA GPS H\,\textsc{i} data cube shown using a linear transfer function on the raw data (left) and the normalising technique described in the text (right). While the raw data preserves a sense of the relative intensity of the data in each channel, the normalised data shows the structures present far more clearly. A rotating movie of the right panel is available online at this url : \href{https://www.youtube.com/watch?v=7r0UJGzLK8M}{https://www.youtube.com/watch?v=7r0UJGzLK8M}.}
\label{fig:scaling}
\end{center}
\end{figure*}

The solution available in \textsc{frelled} is inspired by \textit{ds9}'s ability to automatically vary the transfer function based on the currently displayed channel. The data range displayed in one channel might be, say, 10-15 K, but in another it might be 100-150 K. In this way structures in channels with different dynamic ranges can be shown with equal contrast. I implement this in \textsc{frelled} by having the user choose an option which generates a separate, ``normalised'' cube. The user specifies what fraction of the data \textit{in each channel} should be rendered opaque and what fraction should be transparent. Data values in between the specified minimum and maximum are scaled linearly in opacity. This solution is appropriate for displaying structures, but loses information about the relative flux levels. Figure \ref{fig:scaling} shows a comparison between using the raw and normalised data. 

Rescaling the data either to S/N levels or according to peak flux per channel render the values unphysical, but may be the most appropriate display option. Therefore, \textsc{frelled} allows the user to display one file while performing any quantitative analysis functions (see section \ref{sec:astrofrelled}) on another.

\subsection{Multi-volume display and comparative visualisation}
\label{sec:multi}
\begin{figure}[t]
\begin{center}
\includegraphics[width=84mm]{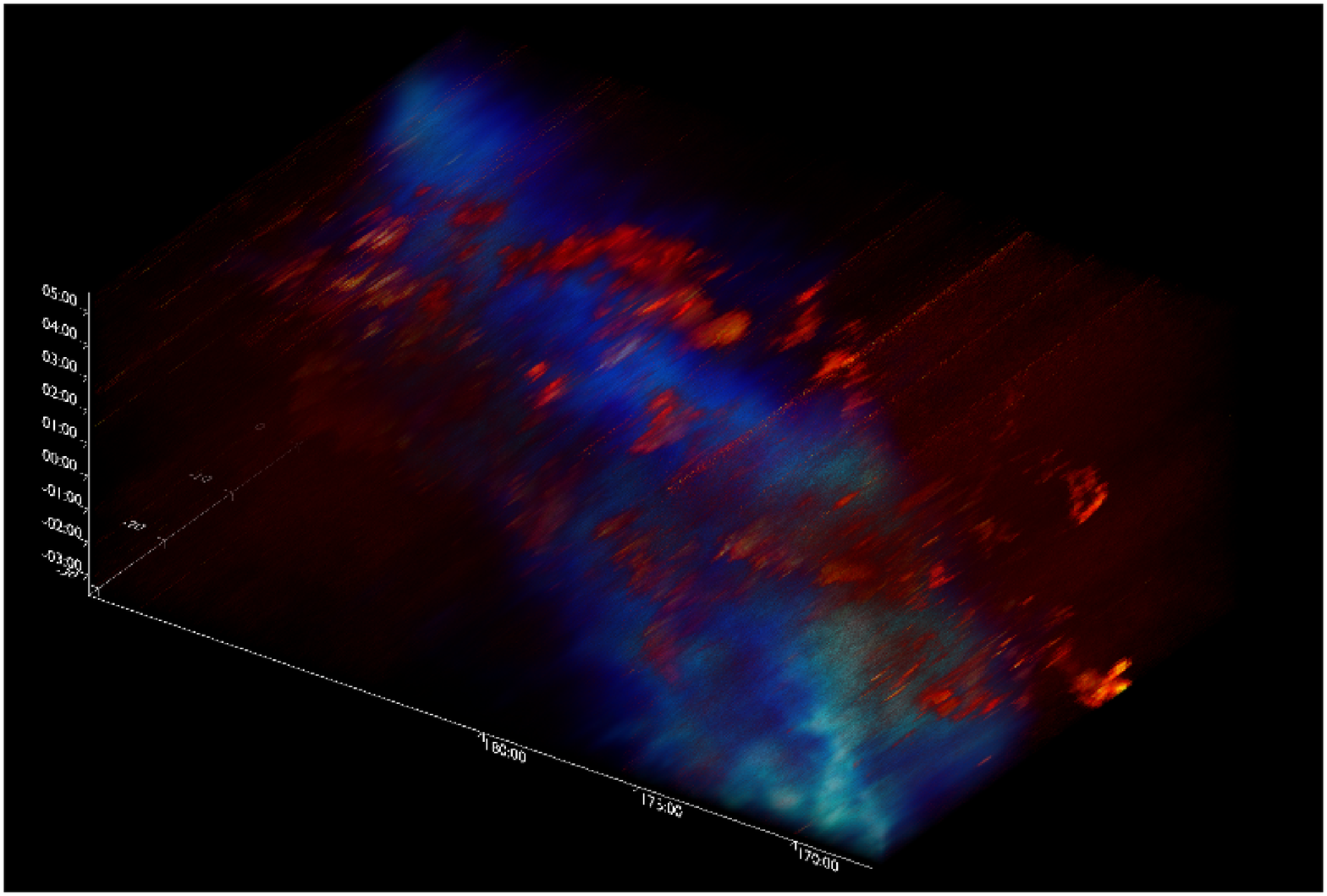}
\caption[hico]{Multi-volume rendering in \textsc{frelled}. Two precisely aligned data cubes (GALFA-H\,\textsc{i} survey in blue, \citealt{peek}; FCRAO CO survey in red, \citealt{heyer}) are shown overlaid in 3D. The display method is to use alternating image planes to display the different data sets. A rotating movie of this figure can be seen online at this url : \href{https://www.youtube.com/watch?v=WQojSoWnpzs}{https://www.youtube.com/watch?v=WQojSoWnpzs}.}
\label{fig:multirender}
\end{center}
\end{figure}

P15 note that the ability to display multiple data sets is an important tool for comparative visualisation. None of the viewers they discuss, nor those I summarise in section \ref{sec:others}, support multi-volume rendering. This feature is implemented in \textsc{frelled} as of version 4.2, as demonstrated in figure \ref{fig:multirender}. Two different data sets can be rendered with different colour schemes and transparency levels (the latter can be altered interactively for each component separately). Each data set is displayed on alternative image planes; thus this causes some penalty in the frame rate. There are many possible uses for such a feature, for example, displaying multi-wavelength data (as in the figure), multi-polarisation data, multi-phase media in numerical simulations, or even arbitrary selections within single-component data sets (for example emission above a specified intensity threshold could be displayed with a different colour scheme to fainter emission).

True multi-volume rendering is at present restricted to two data sets. A simpler alternative of ``blinking'' between data sets is also implemented which has no restriction on the number of components which can be displayed.

I mentioned in section \ref{sec:displaycode} that Blender can display different ``layers'' or workspaces. In fact Blender files can contain many different \textit{Scenes}, each of which contain 20 different layers. A scene is essentially a file within a file, and each can contain completely separate objects, or those objects can be linked across scenes (see section \ref{sec:2D}). Crucially, the viewpoint in each scene is always the same. This means that different cubes can be displayed in different scenes. The images textures of the cubes in different scenes remain in GPU memory,  which means that switching between scenes is very fast. The time to switch the view from one data set to another depends on the number of Blender objects in each scene so is only weakly dependent on the size of the data sets; typically it takes about 1 second to switch between displays.

Blender's mesh objects also make effective tools for comparative analysis. I discuss these in detail in section \ref{sec:sourcefinding}. Here I only note that while it is possible to use mesh objects to generate source catalogues, the reverse is also true - \textsc{frelled} includes tools to import catalogues generated externally. Two special cases of this are the ability to import n-body particles and vectors, allowing comparisons (say) of observational and simulated data. N-bodies are displayed as vertices (single pixel points) while vectors are displayed as cones. The visibility of the particles/vectors is affected by the FITS file : if the opacity is high, particles behind (from the user's perspective) the FITS data will not be seen. There is an `x-ray' option in Blender which forces the meshes to always be visible regardless of objects in the user's line of sight, though at present this cannot be accessed via the \textsc{frelled} menus.

At present it is the responsibility of the user to ensure that the various data sets loaded have the same coordinate systems. Particle data must have the same pixel coordinates as the FITS file. Multiple FITS files should have identical world coordinate systems, being gridded with the same number of pixels and at the same pixel scale.

\subsection{Benchmarks}
\label{sec:bench}
\textsc{Frelled} is designed primarily as a realtime 3D viewer. While the ray-tracing capabilities of Blender can be used to record rotation movies (or time series for simulations), this is far too slow for an interactive view. I test \textsc{frelled} on an HP Elite 7500 series desktop with an Intel i7-3770 quad core 3.4 GHz CPU, 16 GB RAM and a 4GB NVIDIA GeForce GT 640 GPU, and an older Sony Vaio F11Z1E with an Intel i7-720QM quad core 1.7 GHz CPU, 8 GB RAM and a 1GB Nvidia Geforce GT 330M GPU. The HP Elite runs Debian 3.2.51-1 (64 bit) while the Sony Vaio runs Windows 7 (64 bit but with 32 bit Blender and Python). In table \ref{tab:bench}, I compare the loading times, reloading times (see below), display frame rate, and rendering speed for a large (1023$\times$781$\times$241, equivalent to 577$^{3}$ voxels) data cube, shown in figure \ref{fig:scaling} from the VLA Galactic Plane Survey (\citealt{stil}) and a smaller (128$^{3}$ voxels) cube (private FLASH simulation data - see figure \ref{fig:display}). The Sony Vaio lacks sufficient memory to load the larger cube.

\begin{table*}
\tiny
\begin{center}
\caption[bench]{Performance benchmarks for using two different cubes and two different machines. Columns are as follows : (1) Machine used - see text; (2) Size of cube in voxels$^{3}$; (3) Number of projections used; (4) Number of different maps used to control transparency and/or colour; (5) Time to convert and load the data into \textsc{frelled} for the first time in MM:SS; (6) Time to re-open the file in MM:SS; (7) Display frame rate (for each projection if multiple projections used); (8) Time to render a frame (MM:SS) at 500$\times$500 pixel resolution using 7 threads and no oversampling.}
\label{tab:bench}
\begin{tabular}{c c c c l l l c}
\hline
  \multicolumn{1}{c}{(1)Machine} &
  \multicolumn{1}{c}{(2)Cube size} &
  \multicolumn{1}{c}{(3)No.projections} &
  \multicolumn{1}{c}{(4)No.maps} &
  \multicolumn{1}{c}{(5)Loading time} &
  \multicolumn{1}{c}{(6)Reloading time} &
  \multicolumn{1}{c}{(7)Display frame rate} &
  \multicolumn{1}{c}{(8)Frame render time} \\
\hline
  HP Elite & 577 & 1 & 1 & 00:52 & 00:15 & $>$25 & 00:17\\
  HP Elite & 577 & 1 & 2 & 01:19 & 00:29 & $>$25 & 00:20\\
  HP Elite & 577 & 2 & 1 & 01:43+00:18 & 00:15+00:18 & $>$25 / 20 & 02:46\\
  HP Elite & 577 & 2 & 2 & 02:43+00:30 & 00:30+00:28 & 20  / 12 & 03:54\\
  HP Elite & 577 & 3 & 1 & 04:43+00:17+00:22 & 00:15+00:18+00:21 & 20  / 12 / 9 & 03:49\\
  HP Elite & 577 & 3 & 2 & 06:27+00:30+00:35 & 00:30+00:30+00:35 & 15  / 9  / 7 & 03:54\\
  HP Elite & 128 & 1 & 1 & 00:13 & 00:02 & $>$25 & 00:05\\
  HP Elite & 128 & 1 & 2 & 00:16 & 00:02 & $>$25 & 00:07\\
  HP Elite & 128 & 2 & 1 & 00:22+00:01 & 00:02+00:01 & $>$25 / $>$25 & 00:06\\
  HP Elite & 128 & 2 & 2 & 00:28+00:02 & 00:02+00:02 & $>$25 / $>$25 & 00:07\\
  HP Elite & 128 & 3 & 1 & 00:29+00:01+00:01 & 00:02+00:01 & $>$25 / $>$25 / $>$25 & 00:06\\
  HP Elite & 128 & 3 & 2 & 00:37+00:02+00:02 & 00:02+00:02+00:02 & $>$25 / $>$25 / $>$25 & 00:07\\
  Sony Vaio & 128 & 1 & 1 & 00:40 & 00:02 & $>$25 & 00:15\\
  Sony Vaio & 128 & 1 & 2 & 00:49 & 00:02 & $>$25 & 00:15\\
  Sony Vaio & 128 & 2 & 1 & 01:05+00:01 & 00:02+00:01 & $>$25 / $>$25 & 00:14\\
  Sony Vaio & 128 & 2 & 2 & 01:31+00:02 & 00:02+00:02 & $>$25 / $>$25 & 00:15\\
  Sony Vaio & 128 & 3 & 1 & 01:36+00:01+00:01 & 00:02+00:01+00:01 & $>$25 / $>$25 / $>$25 & 00:14\\
  Sony Vaio & 128 & 3 & 2 & 02:15+00:02+00:02 & 00:02+00:02+00:02 & $>$25 / $>$25 / $>$25 & 00:15\\
\hline
\end{tabular}
\end{center}
\end{table*}

Although once the data is loaded the display frame rate is high, initially converting that data into the Blender-readable PNG images and loading them into Blender is much slower than loading in a cube in, say, \textit{kvis}. On the HP Elite, the larger cube took a mere 1 second to load in \textit{kvis}. \textit{Kvis}, which was designed from the ground up as a FITS viewer, is far superior in this regard to \textsc{frelled}, which is essentially a hack to load FITS data into a program never designed with astronomers in mind. I discuss the consequences of this in section \ref{sec:sourcefinding}.

As described above, since \textsc{frelled}'s loading times are so much greater than for \textit{kvis}, it is not necessary to load in all three projections of the data cube - the user can select only one or two, if they wish to save time and/or memory. I account for this in the benchmarks table. The ``loading time'' column shows the time to first load in the data (i.e. convert the cube into PNG images and then display one projection) plus the extra time needed when switching to a different projection. Note that this delay \textit{only occurs once} as the images must be loaded into GPU memory when the user switches the view to a different projection for the first time - after that there is no delay in rotating the view from one projection to any other. Once the user has converted the data with a satisfactory transfer function and altered the display parameters appropriately, they can then save the .blend file, which preserves these display settings. The ``reloading time'' refers to the time needed to re-open this file, which does not require converting the FITS data into PNG images.

Both the time to render an image (using Blender's internal raytracing engine) and the realtime display frame rate  depend (by a factor of a few) on the data being displayed and the orientation of the view. Moving the viewpoint closer to the data causes a drop in frame rate and an increase in render time. For the benchmark table I chose a viewpoint in all cases where the whole of the data was just about visible - zooming in reduces the frame rate to as low as 5 f.p.s., but moving slightly further out increases this to 20 f.p.s. even in the the slowest case. Since the larger cube is asymmetrical, the benchmark table reflects this in the non-linear difference in times for the different projections. I only give the render times for the longest projection in each case.

I chose to render a 500$\times$500 pixel image without oversampling using 7 threads\footnote{Blender uses the GPU for the realtime display but the CPU for rendering.}, with a field of view equivalent to half the extent of the data. It is an unfortunate necessity to render the images rather than simply record screenshots, since this is very much slower. At 3 minutes per frame, a 10 second animation (at standard 25 f.p.s.) would take 12.5 hours to render. While Blender does have the ability to record screenshots, this feature does not function correctly in 2.49 and not usable for our purposes (see section \ref{sec:future}).

Note that all of the above benchmarks are for the standard 3D display mode. In 2D mode, in which only one slice of the data is displayed (and loaded into GPU memory) at a time, loading and rendering speeds improve dramatically. For the larger cube on the HP Elite, using all three projections (but only one colour map since transparency is not necessary in the 2D view), loading speed actually increases to 7 min 33 seconds (but see below). However, reloading speed is slashed to a mere 1 second, competitive with \textit{kvis}. Rendering speed is reduced to 0.25 seconds per frame (the display frame rate remains above 25 f.p.s for the largest cube shown at any angle and zoom level).

Since \textsc{frelled} preserves the original FITS file and generates PNG images, this has the disadvantage of generating a (potentially) large amount of extra data. For the 1.5 GB VLA FITS file we use for benchmarking (see section \ref{sec:bench}), an additional 1.7 GB of PNG data was generated. However, this is a worst-case scenario in which two maps were used for all three projections on a very large cube. For the 16 MB simulation data cube, generating one map for a single projection resulted in an extra 1 MB of data, while generating two maps for all three projections created 6 MB of extra data. Users should therefore be aware that \textsc{frelled} can, typically, approximately double the file space required.

\subsection{Optimisation}
\label{sec:opt}
Conversion of the FITS file is parallelised using Python's ``multiprocessing'' module. This is a trivial operation since each slice of the data can be treated independently.

\begin{figure}[t!]
\begin{center}
\includegraphics[width=84mm]{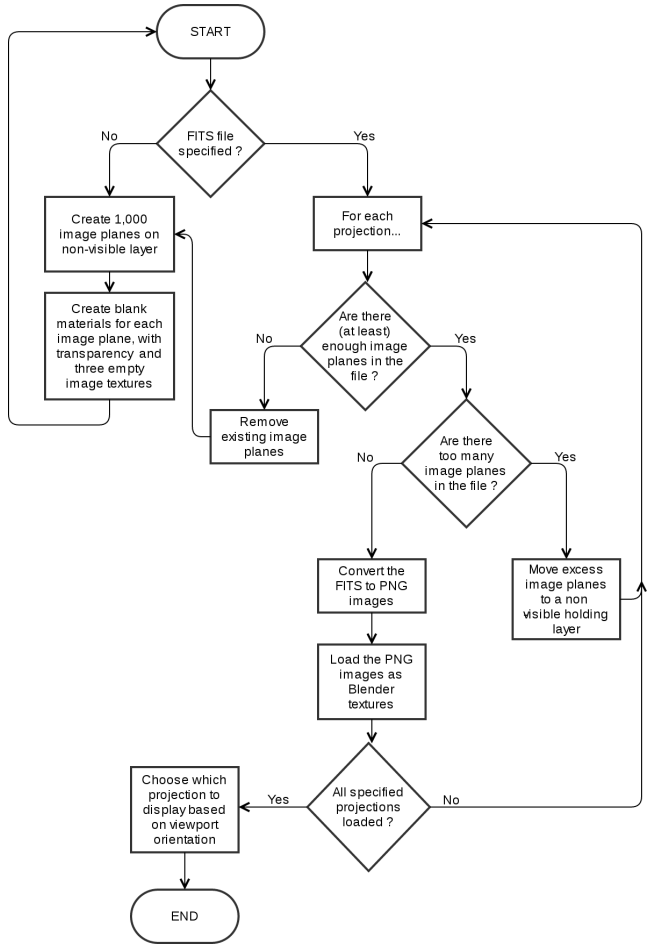}
\caption[Code]{Slightly simplified pseudocode for the \textsc{frelled} script to import FITS data into Blender. If a FITS file is found, the axes of the cube are drawn before any other steps (if existing axes are found these are first deleted). For 2D mode, the blank image planes are animated so as to be on a visible layer when the Blender frame counter is equal to their channel (or slice) number. This simply means that only one slice is visible at a time.}
\label{fig:code}
\end{center}
\end{figure}

Optimising Blender to load the PNG images as rapidly as possible is more complicated (see figure \ref{fig:code} for pseudocode). Creating an object in Blender, for instance, is much slower than transforming an existing object. Older versions of \textsc{frelled} simply created the correct number of planes to hold the images depending on the shape of the data cube. More recent versions (4.0 onwards) contain a pre-existing set of image planes, 1,000 for each projection in both 2D and 3D modes (with appropriate materials assigned in each case).

If a data set has more than 1,000 channels (or spatial pixels along any projection), it is necessary to remove the existing image planes for that projection and re-create the correct number. This step causes a significant slowdown. For example, truncating the large VLA GPS cube to have 1,000 pixels in R.A. (instead of 1,023) reduces the loading speed to 4 min 6 seconds (compared to 6 min 47 seconds for the full cube). 

A similar issue affects 2D mode. In 2D mode, the image planes are animated to move between the visible layer and a non-visible layer depending on the current frame. For example, when displaying channel maps, Blender's frame counter corresponds to the channel number. As in 3D mode, a set of 1,000 image planes per projection are provided in the default file, the difference being that these have also been set up with the necessary animation data. In 2D mode new planes must be both created and animated (a significantly slower extra step) if the image has more than 1,000 pixels along a dimension. Using the truncated cube as above, loading in the data in 2D mode takes only 1 min 58 seconds, compared to 7 min 33 seconds for the full cube.

I estimate that parallelising the code and providing pre-loaded image planes has increased the loading speed of \textsc{frelled} by a factor of approximately 7 in 3D mode and up to 9 in 2D mode, using \textsc{frelled} v3.8 and 4.1 to display the VLA GPS cube described in section \ref{sec:bench} on the quad-core HP Elite.

\subsection{3D display capabilities of other astronomical FITS viewers}
\label{sec:others}
P15 describe in detail four viewers designed explicitly for viewing data in 3D. However, none of these viewers was designed specifically for astronomy - all, for instance, lack the ability to use world coordinate systems, which is a severe handicap (as will become apparent in section \ref{sec:astrofrelled}). Here I briefly compare the 3D display capabilities of \textsc{frelled} with those of other programs designed for astronomy.

\subsubsection{Xray}
\textit{Xray} is part of the \textsc{karma} package. Though now about twenty years old (\citealt{xray}) its loading times are far superior to \textsc{frelled} ($<$ 10 seconds for the VLA GPS cube). Its rendering speeds are considerably lower than \textsc{frelled} ($\sim$ 1 f.p.s.) and the user cannot freely change the angle of the viewpoint - after setting the pitch, yaw and roll via sliders, they must explicitly tell the program to recompute the view. \textit{Xray} is only a viewer and provides no tools to analyse the data. While it does have a 3D cursor, on our HP Elite this feature invariably caused the program to crash.

\textit{Xray} allows the user to interactively define the transfer function and choose different display integrations (e.g. the minimum or maximum voxel along the line of sight). The user is limited to setting the data range via the range selector buttons (e.g. 95\%, 99\%, 99.9\%) - they cannot set precise numerical values. Conversely, the angle of the viewpoint can \textit{only} be set via precise sliders, and not by freely rotating the view (as in \textsc{frelled}) using the mouse. This makes it much more difficult to ``home in'' on a particular feature of interest. However, it has a very easy to use interface for creating animations for presentations (which \textsc{frelled} deliberately emulates). It does not have any facility to stretch data, so if a data cube has a relatively short velocity axis, \textit{xray} is of little use (recall figure \ref{fig:stretch}).

In short, \textit{xray} is useful for quick inspection of data, but cannot be used for detailed analysis.

\subsubsection{Ds9}
\textit{Ds9} is a widely-used, general purpose FITS viewer. It was first released in 1999 and has had 3D display capabilities since 2010. It is, indisputably, an extremely powerful and versatile tool, able to interface with external packages such as \textsc{iraf} and, for example, perform surface brightness profile fitting, aperture photometry, and the result of operations such as smoothing and filtering can be saved directly to the FITS file. It can also be used to access other programs such as \textsc{topcat} and \textit{Aladin}, and, for instance, automatically retrieve and overlay source catalogues. Like \textit{xray}, the transfer function can be controlled interactively, and in 3D the user can choose which line-of-sight integration method to use.

The advantage of \textit{ds9}'s 3D view is that performance does not depend on system configuration. Unlike \textit{xray}, although the user is limited to using sliders to control the viewpoint orientation, the view is updated while they adjust the sliders, rather than waiting until the user forces a redraw . However this update is not instantaneous as in \textsc{frelled} (while updating, only part of the data in the field of view is shown), but depends heavily on the exact display. While the frame rate may drop (at worst) to a few frames per second in \textsc{frelled} with a small field of view, in ds9 this can decrease to many seconds or tens of seconds per frame (or even minutes per frame for a large cube); during the rendering process the user sees a `patchwork' - the data which has been processed is shown with the remaining areas white until the calculations are complete. As with \textit{xray} there is no way to alter the z-axis scale, so kinematic information is hard to discern in cubes with a short velocity axis. All this makes it very difficult to use \textit{ds9} to really explore a data set in 3D, except for very small cubes.

The analysis features in 3D are also very limited. While there is a 3D crosshair, it is not obvious how to use the display to place this in a controlled way, and there is no true 3D cursor (see section \ref{sec:astrofrelled}). It is possible to define 2D regions in the 3D view, but there are no true 3D region capabilities.

As with \textit{xray}, \textit{ds9} provides a quick, simple way to view data in 3D, but not much more than that.

\subsubsection{\textsc{Gaia}}
\textsc{Gaia}, the Graphical Astronomy and Image Analysis tool, offers a similar (if not greater) level of versatility to \textit{ds9}, including photometry, source finding, and interaction with Virtual Observatory tools. However, like \textit{ds9}, its 3D capabilities are limited.

\textsc{Gaia} provides two types of 3D rendering : volume rendering and isosurfaces. Isosurfaces can be useful, for instance, in understanding the structures of sources of very complex 3D geometry by reducing them to simpler information, i.e. the levels of constant flux value. \textsc{Frelled} does not yet implement isosurfaces (but see section \ref{sec:renzo}) as the main motivation for developing the visualisation routine was source extraction. Isosurfaces are not ideal for this purpose, since they intrinsically rely on clipping much of the data so risk the user missing faint sources.

\textsc{Gaia}'s volume renderer has some advantages over that of \textit{ds9}. When panning, zooming, or rotating the view, one does not entirely lose the display but it is replaced with a low-resolution ``preview'' image (rendered at a very high frame rate), which is then recalculated at full resolution once the operation is complete. The user does this not through sliders (which are not present) but through the mouse, making it possible - in principle - to really explore the data set. Unfortunately in practise the navigation interface is very imprecise and centering the view on specific features is extremely difficult. Navigation is also possible via the keyboard (though this is no more precise) in a similar way to \textsc{frelled} (see below), though no GUI buttons are present to complement this. 

Strong artifacts can be visible when altering the view (and even when the display generation is complete). Like \textit{ds9}, and the rendering frame rate depends on the size of the image on the screen. For large data sets at full HD resolution, this can be measured in frames per minute rather than frames per second. The view is also lost if the user switches workspaces and it must be recalculated when they switch back.

\textsc{Gaia} does allow the user to very easily generate 3D axes around the data cube (which assist in orientation and navigation), unlike \textit{ds9}. While \textsc{gaia} does have the capability to load mask polyhedra from an external file, it has no capability of generating or manipulating these interactively.

\subsection{The advantages of Blender's interface}
From the above discussion it is apparent that existing astronomical viewers have only rudimentary 3D display capabilities. Two features in particular are lacking : 1) The ability to maintain the full detail of the data in the field view at any screen resolution and zoom level, especially while altering the viewpoint, while also maintaining a high frame rate (as discussed above, other viewers suffer from displaying a patchwork view of the data while updating the view and/or other artifacts); 2) The ability to intuitively and precisely navigate through 3D data.

By using Blender, \textsc{frelled} completely avoids these problems. In Blender the frame rate in the realtime display (see section \ref{sec:bench}) is not strongly affected by screen resolution or zoom level, nor is it necessary to reduce the level of detail when altering the viewpoint angle. The extremely slow `realtime' display of \textit{ds9} (at least for large data sets) means that the user is not really viewing an interactive 3D display at all, but simply changing from one fixed viewpoint to another. The experience when freely changing the view without loss of frame rate or detail is wholly different.

While art is normally described as an intuitive process, artists require tools which are both intuitive and precise. The navigation interface in Blender is far in advance of those of any of the tools discussed above. Blender has a true 3D cursor which is easy to place (see section \ref{sec:sourcefinding}) either using the mouse or by exact numerical values, and the user can then instantly re-orient the view on this position. This can also be used as the pivot point around which the view rotates. \textsc{Frelled} provides altitude/azimuth sliders to control the viewpoint, if necessary. Translating the view can be done with very high precision in Blender - unlike in \textsc{gaia} the movement of the view correlates perfectly with the mouse movement. Navigation is also possible via the keypad, and \textsc{frelled} provides GUI buttons to support these options. For really exploring a 3D data set, a straightforward navigation interface which makes it easy to examine particular features is not an optional extra : it is an essential, fundamental part of the process.

\section{Using \textsc{frelled} in astronomy}
\label{sec:astrofrelled}
\textsc{Frelled} was originally designed specifically for source-finding in H\,\textsc{i} data cubes. It has since been adapted to work as a viewer for simulation data, though most of its analysis tools are geared towards H\,\textsc{i}. While the goal is to provide a general-purpose astronomical viewer, at the present time some features are still hard-coded to assume the data is H\,\textsc{i} (in particular drawing the velocity axis).

\subsection{Source finding}
\label{sec:sourcefinding}
As discussed in \ref{sec:displaycode}, source extraction demands using minimal (or no) clipping of the data. I have already described in detail how this is accomplished in \textsc{frelled}. In T14, we demonstrated that using \textsc{frelled} gave the same level of reliability of source extraction as using an established viewer such as \textit{kvis}. Describing a search of a 330$\times$270$\times$4096 voxel cube containing 334 sources :
``One of us (HH) searched the cube using KVIS, while another (RT) searched with our new viewer. HH found 20 sources that RT did not, and RT found 18 sources that HH did not. Such differences are inevitable given the subjective nature of the procedure; however, RT was able to complete the process in a day, whereas HH required approximately six weeks.'' We also note that HH did not mask the sources, as this is very difficult to do in \textit{kvis} (see below).

The key to this dramatic increase in visual source finding speed is the ability of the user to interactively mask regions of the data - especially in combination with the realtime 3D view of the data. The user can place Blender's 3D cursor using the mouse. It is straightforward to define the exact position of the cursor since its position is restricted to the current plane of the screen. For example if the user is viewing the XY projection (sky coordinates), they can only alter the right ascension and declination of the cursor, not its redshift. Similarly in the XZ projection they can only alter its right ascension and redshift, not its declination. Switching back and forth between the projections (which is instant) enables the cursor to be placed very precisely (numerical values can also be specified).

Once the user has placed the cursor at the center of the source, they can then add a Blender object using the \textsc{frelled} script. By default this object will be a black cuboid, which they can scale along any and all axes using the keyboard and mouse (again it is also possible to enter precise numerical values). With practise this is an extremely fast procedure that takes no more than a few seconds per source (note that I am primarily using \textsc{frelled} to catalogue single-dish data where most sources are unresolved). It is easier to do this in 3D since the user can see how much of the source they have masked along all three axes at once - there is not much need to change the projection once the source center has been determined. This method reduces the limiting factor on the speed of visual source extraction, which has been \textit{recording} sources rather than actually \textit{finding} them (see below for details).

It is possible to mask data in a similar way using a combination of the \textsc{miriad} task \textit{immask} and \textit{kvis}. However, this is scarcely a viable alternative. Firstly, it is not interactive. The user must determine the precise numerical values for the minimum and maximum coordinates of the box and record them by hand. They must then use \textit{immask} to apply the mask, re-load the cube in \textit{kvis} and navigate back to the source in question. If a mistake is made, the mask file must be deleted and a new one created - essentially all the masks must be re-applied. Moreover, this mask functions only as a mask, and nothing else.

\begin{figure}[t]
\begin{center}
\includegraphics[width=84mm]{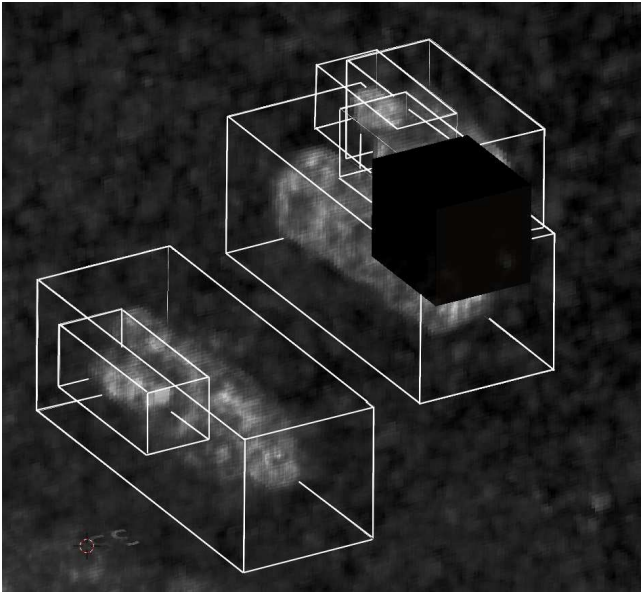}
\caption[masks]{Masking data in \textsc{FRELLED}, showing wireframe and solid display modes. These modes can be switched interactively. Using Blender's mesh modelling tools, it is possible to create masks of arbitrarily complicated geometry.}
\label{fig:masks}
\end{center}
\end{figure}

By allowing the user to interactively create an object which is used as a mask, \textsc{frelled} essentially eliminates all of these difficulties. As well as being much faster to create (see below), the masks can be moved interactively, which is not possible using \textit{kvis}. They can also be toggled so that they are displayed either as true masks (solid black in place of the data) or transparent wireframes. An example of mask objects is shown in figure \ref{fig:masks}. Furthermore, since \textsc{frelled} can use world coordinates\footnote{The relevant FITS header data is written to a human-readable ASCII file, which \textsc{frelled} accesses as required.}, it is straightforward to generate a source catalogue and/or compare detections with other data on the fly (see below). The user rarely even needs to know the coordinates of the source they have found.

While \textsc{frelled}'s file loading times are mostly abysmal compared to conventional viewers such as \textit{ds9} or \textit{kvis}, for source finding this is irrelevant. For the sake of a few extra minutes of loading time, weeks of source extraction time can be avoided. 

However, it is important to note that this is only true provided the source density is sufficiently high. It is interesting to ask : at what point does \textsc{frelled}'s high loading time offset its fast masking capabilities ? This will depend on the size of the cube and system configuration. As a ballpark estimate, I masked ten random sources in a standard AGES subset cube (330$\times$270$\times$925 voxels). The cube took approximately 1 min 30 seconds to load on our HP Elite and it required an average of 30 seconds to mask each source. In comparison, in \textit{kvis} is took less than 1 second to open the cube, but about 1 min 50 seconds to mask each source, on average. Therefore it would still have been faster to use \textsc{frelled} for this cube if there were only two sources present. Even with the (truncated) 577$^{3}$ voxels cube described in sections \ref{sec:bench} and \ref{sec:opt} (which is about the largest cube \textsc{frelled} is capable of displaying anyway, on a contemporary desktop machine), using \textit{kvis} would only have been faster then \textsc{frelled} if no more than two sources were present.

Note that in the case of a real source-finding process, \textit{kvis} can be significantly slower than in this simple test. The user must record source coordinates, which must also be done by reading values from the screen. Masking sources can be much slower than in the ideal case if source density is high, as it can be difficult to determine where one source ends and other begins. Applying the masks should be done whenever a single source is found, rather than after finding ten sources as in the test, otherwise there is a risk of cataloguing sources twice. If the user chooses to avoid masking the sources, they must continually check their catalogues to see if they have found a source or not. All of these factors mean that the factor of four difference in masking speeds in our idealised test translates into a very much larger difference in practise.

\subsection{2D mode}
\label{sec:2D}
P15 describe the necessity of having linked 3D and 2D views. As we are limited by Blender's display to only displaying the sum of values along the line of sight (and not minimum or maximum as most other viewers), the faintest sources are difficult to see in 3D mode. Noise is rarely the ideal Gaussian with a peak at exactly zero, and structures in the noise can overwhelm faint sources even with different line-of-sight display options. Some data, however little, must usually be clipped to give a useful 3D display. In a 2D display, on the other hand, no data need be clipped at all, and structures in the noise do not necessarily obstruct the view of fainter sources. For these reasons, a 2D display mode remains a valuable tool for source finding.

Objects created in the 3D view are automatically linked to the 2D display. Any changes made to an object in one view are propagated to the other. Masking sources is slightly more difficult in the 2D view since the user has to change the visible channel (or slice) and then adjust the size of the mask region (as opposed to the 3D view where it is not necessary to change the viewpoint). Therefore a sensible approach is for the user to first mask all the sources visible in the 3D view (i.e. the bright ones) and then search the cube in 2D to find the fainter sources.

\subsection{Other features enabled by interactive mask objects}
\subsubsection{What's in a name ?}
It is often the case that H\,\textsc{i} detections do not appear especially interesting in the H\,\textsc{i} data cube itself. For instance, one cannot tell from the H\,\textsc{i} data alone whether a galaxy has a particularly high H\,\textsc{i} mass-to-light ratio or is unusually compact compared to the stellar emission. Such information can only be obtained after a much more detailed analysis. Once determined the user may wish to re-asses the H\,\textsc{i} data in light of this new information and perhaps process it in different ways.

One very simple but indispensable feature of Blender objects is that they are given names, unlike regions in \textit{ds9} or masks in \textit{kvis}. This means that it is very simple to find the objects again later - one does not even need to know their coordinates, just the name (or even just a number depending on the naming scheme used). The viewpoint in \textsc{frelled} can be automatically centred on the object using its name or coordinates.

Blender does not allow objects to have names longer than 21 characters. \textsc{Frelled} assumes objects have the naming format STRING\_XXX, where XXX is a numerical suffix. The string part of the name is limited by \textsc{frelled} to be 13 characters long. Therefore the user can theoretically generate up to 10 million objects.

Through the Python interpreter, it is possible in principle to perform any type of numerical analysis on the specified sources. I briefly describe those features which are currently implemented in \textsc{frelled} below. While all of the following require a region to be defined, it is (if needed) straightforward to define a region as large as the entire data cube. Most operations can also be applied to multiple regions simultaneously.

Note also that the display of the FITS data can be disabled at any time, to ensure a clearer view of the other maps (see below) \textsc{frelled} can generate.

\subsubsection{Contour plots}
The realtime display in \textsc{frelled} provided a qualitative view of the data. More quantitative displays are possible using contour plots. The user has only to specify the minimum, maximum, and interval values. At present the interval size is fixed. Contours are enabled through the \textit{matplotlib} Python module and either displayed as PNG images, or converted into Blender's internal mesh format. This is done through a combination of internal \textsc{frelled} code and a ``solidify'' code provided with Blender, which converts the one-dimensional lines generated in \textit{matplotlib} into 3D structures.

To help produce conventional 2D plots (e.g. for publications), the colours of the contours can be set according to the velocity (or whatever the third axis of the cube happens to be) of the sources. These can be defined by the axis range of the entire cube or by the subset of selected sources. For examples, see T14 appendix B.

If a region spans more than one channel, the contours plotted will be of the integrated flux (moment 0) for that region. The colour assigned will depend on the systemic velocity of the source, i.e. the central channel of each region.

\subsubsection{Renzograms}
\label{sec:renzo}
Renzograms are contour plots of the flux in each channel of a source, with each channel assigned a different colour. The colours are set in the same way as for contour plots : they can either depend on the global velocity axis, or the particular velocity range of the selected regions.

In 3D, renzograms are essentially slices of an isosurface. The ability to generate true isosurfaces will be included in a future version of \textsc{frelled}.

\subsubsection{Moment maps}
Moment maps can show a variety of different quantities, though currently only moment 0 (integrated flux) maps are supported in \textsc{frelled}. They can be useful in increasing sensitivity to extend flux, as discussed in T14. Moment maps in \textsc{frelled} are limited to greyscale images.

\subsubsection{Query NED and the SDSS}
The support for world coordinates means it is straightforward for \textsc{frelled} to query the NASA Extragalactic Database and the Sloan Digital Sky Survey (\citealt{sdss}). This is obviously useful to discover if a source has a counterpart at other wavelengths, which can be important in - for example - deciding if a source requires follow-up observations. Queries are performed using a near-position search in NED and by opening the SDSS navigate tool at the coordinates specified (both by simply opening the NED or SDSS URL in a web browser with the appropriate parameters). No attempt is made by \textsc{frelled} to match the objects found through the query with the source identified in the data - this is a non-trivial operation for which it is better to allow the user full control (for instance, the user may demand only optical redshift data and not H\,\textsc{i} redshifts, or sources found in NED may not have any redshift information in which case the user must decide if a closer or brighter source is more likely to be the actual counterpart; the user might also decide that \textit{none} of the counterparts are plausibly associated with their source).

\subsubsection{Direct SDSS image overlay}
RGB images from the SDSS can be directly displayed in \textsc{frelled}. Their field of view is set to be the same as the region queried. Thus is it is very easy to overlay, for example, H\,\textsc{i} contours on an optical image with a single mouse click.

The SDSS does not permit downloading images larger than 2048$\times$2048 pixels. Images are normally extracted at the maximum resolution of 0.3961$''$. If the region is so large that the image would exceed the maximum limit, \textsc{frelled} adjusts the resolution to keep the field of view the same.

\subsubsection{Calculate total flux and statistics}
It is straightforward to compute the sum of values in a specified region. Since Blender's internal Python enables us to check if a point is inside a mesh of arbitrary complexity, this allows analysis of regions of any shape. \textsc{Frelled} reports both the raw sum of values and the beam-corrected total integrated flux, as described in T14. An array of the locations of points inside the mesh is also written to a text file, allowing the user to write their own scripts to process data from regions of complex shapes. Basic statistics (minimum, maximum, mean, median and standard deviation) of the raw values in the selected region are also calculated.

\subsubsection{A \textsc{miriad} GUI}
\textsc{Miriad} is a widely-used tool in radio astronomy for data reduction and also spectral line analysis (\citealt{miriad}). In particular the \textit{mbspect} package is an accepted standard way in which to determine the precise coordinates of a source and also measure its line width and total flux. This requires the user to numerically specify their initial estimate of the coordinates, the velocity (or frequency) range of the source, any velocity range that should be masked to avoid measuring (for instance) RFI, and the name of a log file to save the extracted spectrum.

\textsc{Frelled} allows the user to access \textit{mbspect} via the user-created regions. These feed the (world) coordinates and velocity range of the source into \textit{mbspect}. Any other regions which are at similar coordinates but at a different velocity are also used to mask data, preventing the baseline fit in \textit{mbspect} from being unduly affected by other sources. \textsc{Frelled} can access \textit{mbspect} in two different ways : 1) Interactively - if a single source is selected, \textit{mbspect} is run without saving any information to a log file, and the fitted spectrum is displayed in an X-Window; 2) Non interactively - if multiple sources are selected, all the necessary parameters are saved to input files, named according to the regions, with log files specified. In both cases, since \textsc{frelled} does not offer direct access to all \textit{mbspect} parameters, the idea is provide a useful first look at the data and reduce the workload when fine-tuning the parameters later.
	
Regions in \textsc{frelled} can also be used to mask data with \textsc{miriad}. The advantage of doing so is that \textsc{miriad}'s \textit{immask} sets values in the specified region to NaN, rather than merely hiding them as in \textsc{frelled}. This can be useful, for example, when analysing the noise characteristics of a cube.

While only a few \textsc{miriad} tasks are current accessible via \textsc{frelled}, there is no reason why any other tasks which benefit from having a specified region (essentially any which analyse individual sources rather than entire data sets) cannot be similarly utilised. \textsc{Frelled} accesses \textsc{miriad} simply by writing the necessary parameters to a text file and running the appropriate task, avoiding the need to re-invent the wheel and recode all of the well-tested \textsc{miriad} features. Thus, extending \textsc{frelled} to use other \textsc{miriad} tasks is a straightforward procedure.

\subsection{Simulations and animations}
Numerical simulations present a few special concerns. They usually contain far more information than observational data : the user will probably opt to display true 3D spatial information rather than two dimensions of space and one of velocity. A \textsc{frelled} tool can display velocity information as 3D mesh arrow vectors. Alternatively the user could use a separate FITS file to display velocity information as colour (see section \ref{sec:displaycode}). There is also support for loading n-body particle data (as with vectors these are loaded as Blender meshes).

One other key difference between numerical simulations and observational data is that simulations usually have multiple timesteps. Both the particle and FITS data can be animated in \textsc{frelled} - in the latter case the user simply supplies a file containing a list of the FITS files they wish to animate. It is also possible to render batches of simulations (currently they should all have similar data ranges, since it is not possible to specify a different transfer function for each simulation).

For presenting data, simple rotation movies are also supported (this can be used in conjunction with time series data). The user specifies the position and orientation of the camera and the number of frames for it to complete one rotation.

\section{Current limitations and future developments}
\label{sec:future}
P15 describe in detail the requirements for an H\,\textsc{i} FITS viewer. I have described many of those features  which \textsc{frelled} already offers throughout this article. Here I describe the limitations of \textsc{frelled}, including features that potentially could be implemented and those which are impossible.

Of the requirements listed in P15, three are likely impossible in \textsc{frelled} and/or Blender, short of involving a Blender developer. Firstly, we cannot interactively change the transfer functions. In most FITS viewers, the user can interactively change the contrast of the displayed image by using the mouse and/or numerically setting the minimum and maximum data values. This is not possible in \textsc{frelled} without re-converting the FITS file into PNG images an reloading them. As I showed in section \ref{sec:bench}, this is prohibitively slow except for small data sets, and cannot be considered to be an interactive feature. While the contrast of the PNG images can be directly controlled within Blender, this does not affect the realtime display and the rendering the result is also prohibitively slow.

Fortunately, \textsc{frelled} is not entirely without the ability to control the image display. Controlling the transfer function is useful since the data often has a high dynamic range. Studying extended, complex bright features requires a different data range to that suitable for even detecting unresolved faint sources. To this end, there are some options in \textsc{frelled}/Blender to interactively enhance the visibility of different features :
\begin{itemize}
\item The maximum opacity of the data - see figure \ref{fig:display}. Making the data more opaque makes fainter sources easier to see, but reduces the visibility of structures within bright features.
\item Clipping of faint features. Blender allows all features below a specified transparency threshold to be rendered completely invisible. This is useful for hiding faint features (e.g. noise) that may obscure the view of other, brighter sources.
\item Adjusting how the colour maps are applied to the data. For instance a texture map can affect only the transparency or the transparency and the colour (which results in a higher contrast display), and these mapping modes can be toggled interactively.
\end{itemize}

Therefore, while \textsc{frelled} does not allow the transfer function to be controlled interactively, I believe it has enough alternative options that this is only a minor deficiency.

More problematically, it is not possible to change the line-of-sight integration method. As mentioned in section \ref{sec:2D} Blender is limited to displaying only the sum of values. Although I have not found this to be a severe handicap in practise, it would certainly be advantageous to be able to display, say, the peak flux instead (as is possible in \textit{xray} and \textit{ds9}). \textsc{Frelled} is restricted by the whims of the Blender developers in this case (or more accurately by the needs of Blender's main user base).

The third feature which is essentially impossible is to interactively smooth the data. As noted in P15, this can help reveal fainter features by increasing sensitivity - but of course this also decreases resolution, hence the need to be able to interactively switch between smoothed and unsmoothed cubes. \textit{Ds9} is able to interactively smooth the data on the fly, a process which is somewhat slow but not prohibitively so (about 38 seconds for our large VLA GPS cube on the HP Elite). In contrast it takes about two minutes to smooth the PNG images for the 128$^{3}$ voxel cube - much longer than the 40 seconds it takes to load the data into \textsc{frelled}. Possible workarounds for this would be the implementation of isosurfaces, which would allow smoothing of much smaller subsets, or using pre-computed smoothed cubes (see section \ref{sec:multi}) - the latter would be slow to initially load in, but switching between different cubes would be instantaneous.
 
P15 also discuss the size of the data sets that various SKA precursor surveys will return. As shown in section \ref{sec:bench}, \textsc{frelled} is certainly not capable of handling the enormous $\sim$4000$^{3}$ voxel data sets such surveys will produce. It is in fact possible to already visualise such enormous data sets, but this requires access to a large cluster of GPUs (\citealt{hass}), something the average user does not have. \textsc{Frelled} is, however, fully capable of handling subset cubes returned by automatic algorithms (see P15) on a standard desktop workstation. For most contemporary data sets, \textsc{frelled} is perfectly adequate.

Most of the other features mentioned in P15 are already partially implemented in \textsc{frelled} and it should be straightforward to develop them fully. It already supports moment 0 maps, extending this to other moments (e.g. velocity, velocity dispersion, peak flux) would be straightforward. Isosurfaces are not yet supported but renzograms are - and as noted in section \ref{sec:renzo}, these are essentially slices of isosurfaces. It is quite literally a matter of ``joining the dots'' to turn these into isosurfaces, for which a Blender script already exists (see \href{http://sourceforge.net/projects/pointcloudskin/}{this url}\footnote{\href{http://sourceforge.net/projects/pointcloudskin/}{http://sourceforge.net/projects/pointcloudskin/}}).

\section{Summary}
\label{sec:summary}
I have described the capabilities of \textsc{frelled}, a FITS viewer designed specifically to display astronomical FITS data cubes in 3D. I have discussed in detail the problems of visual source extraction, noting that \textsc{frelled} is an effective way to alleviate one of its principle difficulties~: the low speed in comparison with automatic programs.

I have demonstrated that using interactive objects to mask and catalogue data, with the appropriate interface (particularly using a 3D display) is a highly effective way to increase the visual source extraction process. This makes it possible to catalogue hundreds of sources per day, which is sufficiently fast that there is little point in \textit{not} undertaking such a search.

I have also discussed how visual source extraction out-performs most algorithms in terms of reliability, if not completeness. It is worth emphasising that both techniques have an important role. In large data sets, the subjective nature of visual source extraction makes it desirable to do multiple searches - both by multiple people and algorithms (who/which may all have different biases).

While visual extraction will always been subjective, at the current time so-called automatic algorithms fare little better in this regard. At low S/N levels (which is often where the most interesting sources are found), their reliability levels are so low that their catalogues require human inspection anyway. Moreover, while it is obviously desirable to have a catalogue which is complete to some (say) flux limit, there is also value in finding every source possible without imposing any rigorous selection criteria. Imposing criteria for cataloguing sources too strictly runs the risk of avoiding important serendipitous discoveries.

Finally, I note that many of the features provided by \textsc{frelled} are, technically, also possible with other software. Regions of data can be masked using a combination of \textsc{miriad} and \textit{kvis}, but far more slowly and laboriously. One can input all of the necessary parameters to analyse a spectral line in \textit{mbspect} by carefully inspecting the data and typing them in by hand. What \textsc{frelled} attempts to do is provide not so much new features but simply a better \textit{interface} to control them. In this way, it becomes possible to do visual source finding and analysis much more rapidly and on a larger scale than was previously possible. Similarly, the far superior navigation interface of Blender makes it possible to explore data sets in a completely different way to other 3D viewers. While \textsc{frelled} itself may not be used in analysing the large data cubes returned by SKA precurosr surveys, I believe that similar features, implemented in other software, would bring tremendous benefits.

\section*{Acknowledgements}
The authors wish to thank Kevin Douglas for graciously providing the multi-wavelength data shown in figure \ref{fig:multirender}. 

This work was supported by the project RVO:67985815 and by the Czech Science Foundation project P209/12/1795.

This work is based on observations collected at Arecibo Observatory. The Arecibo Observatory is operated by SRI International under a cooperative agreement with the National Science Foundation (AST-1100968), and in alliance with Ana G. M\'{e}ndez-Universidad Metropolitana, and the Universities Space Research Association. 

This research has made use of the NASA/IPAC Extragalactic Database (NED) which is operated by the Jet Propulsion Laboratory, California Institute of Technology, under contract with the National Aeronautics and Space Administration.

This work has made use of the SDSS. Funding for the SDSS and SDSS-II has been provided by the Alfred P. Sloan Foundation, the Participating Institutions, the National Science Foundation, the U.S. Department of Energy, the National Aeronautics and Space Administration, the Japanese Monbukagakusho, the Max Planck Society, and the Higher Education Funding Council for England. The SDSS Web Site is http://www.sdss.org/.

The SDSS is managed by the Astrophysical Research Consortium for the Participating Institutions. The Participating Institutions are the American Museum of Natural History, Astrophysical Institute Potsdam, University of Basel, University of Cambridge, Case Western Reserve University, University of Chicago, Drexel University, Fermilab, the Institute for Advanced Study, the Japan Participation Group, Johns Hopkins University, the Joint Institute for Nuclear Astrophysics, the Kavli Institute for Particle Astrophysics and Cosmology, the Korean Scientist Group, the Chinese Academy of Sciences (LAMOST), Los Alamos National Laboratory, the Max-Planck-Institute for Astronomy (MPIA), the Max-Planck-Institute for Astrophysics (MPA), New Mexico State University, Ohio State University, University of Pittsburgh, University of Portsmouth, Princeton University, the United States Naval Observatory, and the University of Washington.

{}

\end{document}